\documentclass[10pt,draft]{dis03}
\usepackage{epsf,amsmath}
\usepackage{epsfig}

\newcommand{\Etg}{$E_T^{\gamma}$ }
\newcommand{\etag}{$\eta^{\gamma}$ }

\begin{document}

\title{PROMPT PHOTON PRODUCTION AT HERA
\thanks{on behalf of the H1 and ZEUS collaborations.
Talk presented at DIS03, St. Petersburg.
}}
\author{RACHID LEMRANI\\
DESY \\
Hamburg, Germany\\
}

\maketitle

\begin{abstract}
\noindent
    Results are presented on the production of
isolated prompt photons in photoproduction (virtuality $Q^2 < 1$ GeV$^2$)
and in deep inelastic scattering ($Q^2 > 35$ GeV$^2$).
    The results are reasonably well described by pQCD
    calculations in next to leading order. Comparisons to the 
    Monte Carlo models PYTHIA and HERWIG are also presented.

\end{abstract}

\section{Introduction}

  The production of isolated prompt photons is studied at HERA
  in photoproduction and in deep inelastic scattering (DIS).
  In  electron\footnote{The term
  ``electron'' is used both for electrons and positrons.} proton scattering,
 photons are emitted
 by the incoming electron and interact with the proton which are 
 quasi real (photoproduction) or have substantial virtuality (DIS).
 These photon-proton interactions
 can lead to the process of so called prompt photon emission
 which is sensitive
 to the partonic substructure of the proton and, in case of resolved 
 photon interactions, also to that of the exchanged photon. 
 It is an advantage of this reaction that an isolated photon at large
  transverse energy \Etg can be related directly to the partonic event
  structure.
 In contrast
   to jet measurements,
   here the partonic structure is not hidden behind the non perturbative
   hadronisation process. Further information on the dynamics
  of the process can be obtained if prompt photons are measured
  together with jets.

  Preliminary results on inclusive prompt photon production
  have previously been presented  by the H1 collaboration~\cite{h1prel}.
  In the present report, photoproduction results
 are presented where in addition
  to the prompt photon a jet is detected.
 These measurements are confronted 
   with NLO calculations using the program of
   Fontannaz, Guillet and Heinrich
~\cite{Fontannaz:2001ek}.
Also presented are first results on prompt photon production in DIS
 by the ZEUS collaboration, which are compared with the
 event generators  PYTHIA~\cite{Sjostrand:2000wi} and HERWIG 
and with pQCD NLO calculations
 by Kramer and Spiesberger based on~\cite{Gehrmann-DeRidder:2000ce}. 
  In DIS the resolved photon contribution is
 suppressed and, in contrast to photoproduction, is not taken into account
 in the calculation.

\section{Experimental Conditions}

 The data were taken in the years 1996-2000
 and correspond to more than 100~pb$^{-1}$ for each
 experiment.

  The prompt photons are identified by dense calorimetric clusters
  without associated tracks.   
  The main experimental difficulty is the separation of the prompt photons 
  from hadronic background, in particular from 
  signals due to $\pi^0$ mesons and, to less extent, from $\eta$'s,
 as for those, at high energies, the
  two energetic decay photons cannot be resolved in the calorimeters.
  The neutral mesons are predominantly produced in jets.
  Therefore an isolation condition is imposed that
 the transverse energy, $E_T^{cone}$, in a cone
 around the $\gamma$ candidate
 in the  plane of pseudorapidity
 and azimuth,
 $\Delta r = (\Delta\eta ^2 + \Delta \phi ^2)^{1/2} < 1$,
 does not exceed
 10\% of the transverse energy \Etg of the prompt photon candidate.  
 Minimal \Etg is 5 GeV for both experiments. 
 After these cuts, the background is still of similar size as
  the prompt photon signal.
  The signal is thus extracted by a combination of
  discriminating shower
  shape functions exploiting the fact the prompt $\gamma$ showers are
  more compact than the $\pi^0$ and $\eta$ showers.

The H1 results are given for $142 < W < 266$ GeV and $Q^2 < 1$ GeV$^2$, 
where $W$ is the $\gamma p$ center of
 mass energy, and $Q^2$ the virtuality of the exchanged photon.
 Associated jets are reconstructed using the inclusive $k_T$
 algorithm~\cite{Ellis:tq}
 with the conditions 
$E_T^{jet} > 4.5$ GeV
 and $-1 < \eta^{jet} < 2.3$ for the jet energy and pseudorapidity
 respectively. 
The ZEUS DIS results are given for  $Q^2 > 35$~GeV$^2$,
 the associated jets are reconstructed using the cone algorithm with
 $\Delta r < 0.7$ and 
$E_T^{jet} > 6$ GeV
 and $-1.5 < \eta^{jet} < 1.8$.

\section{Photoproduction Results}
\begin{figure}[htb] \unitlength 1pt
\vspace*{265pt}   
\begin{center}
\includegraphics{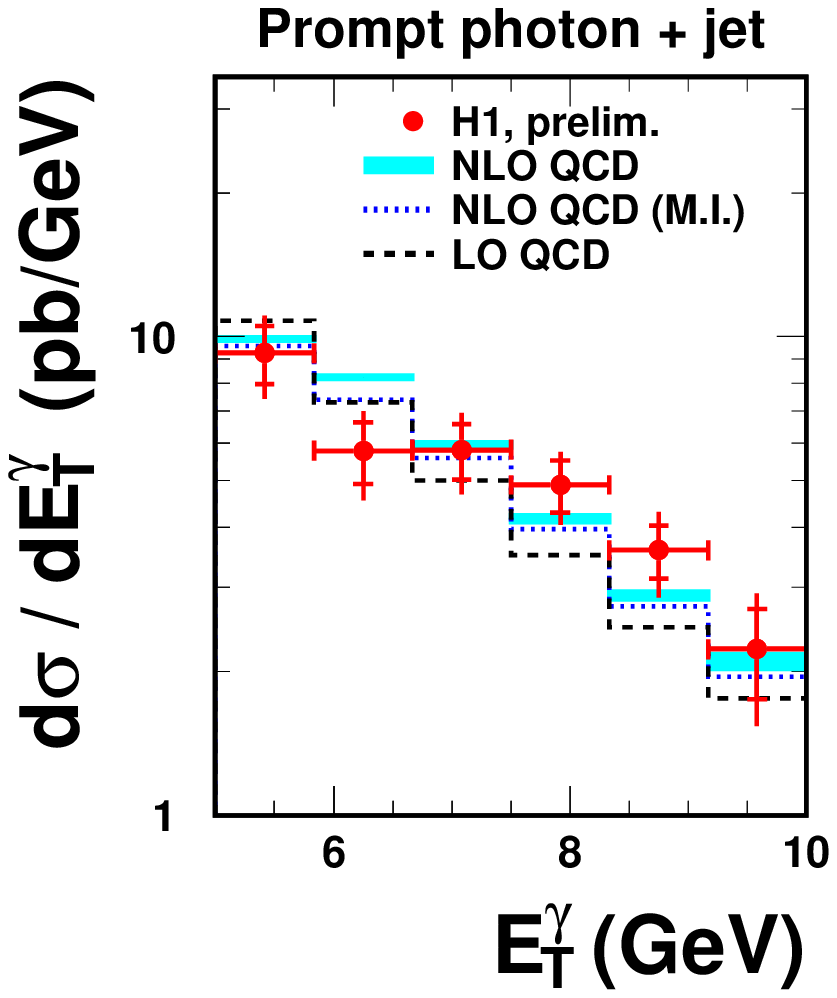}
\includegraphics{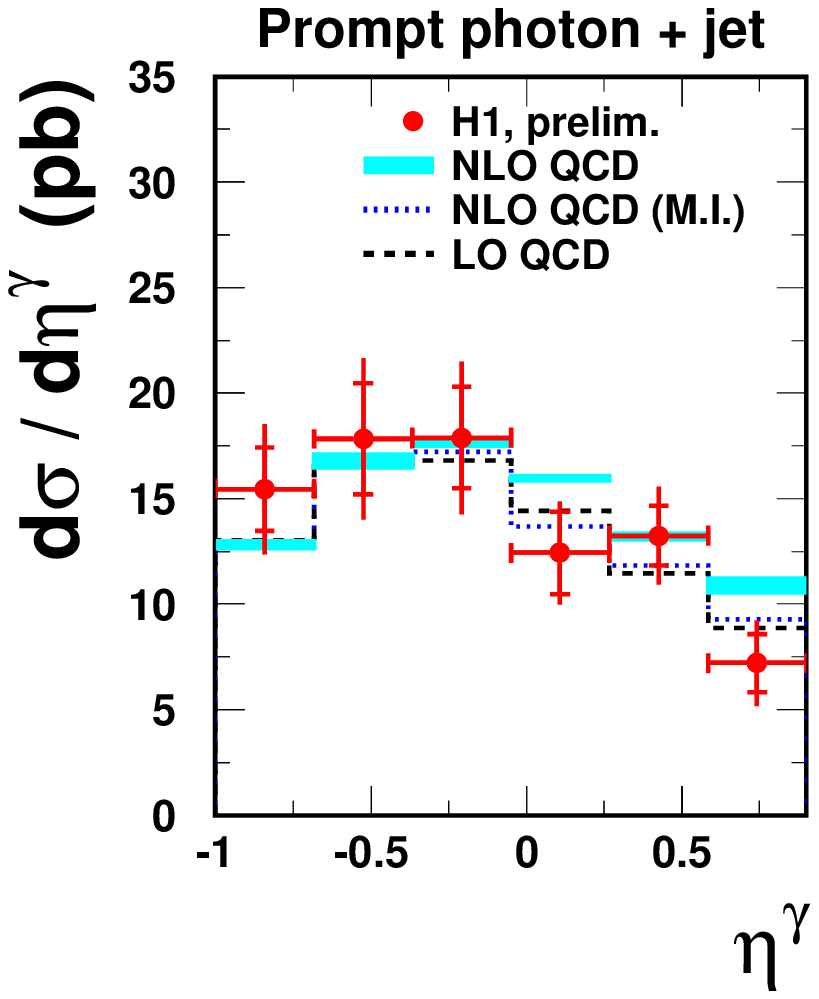}
\includegraphics{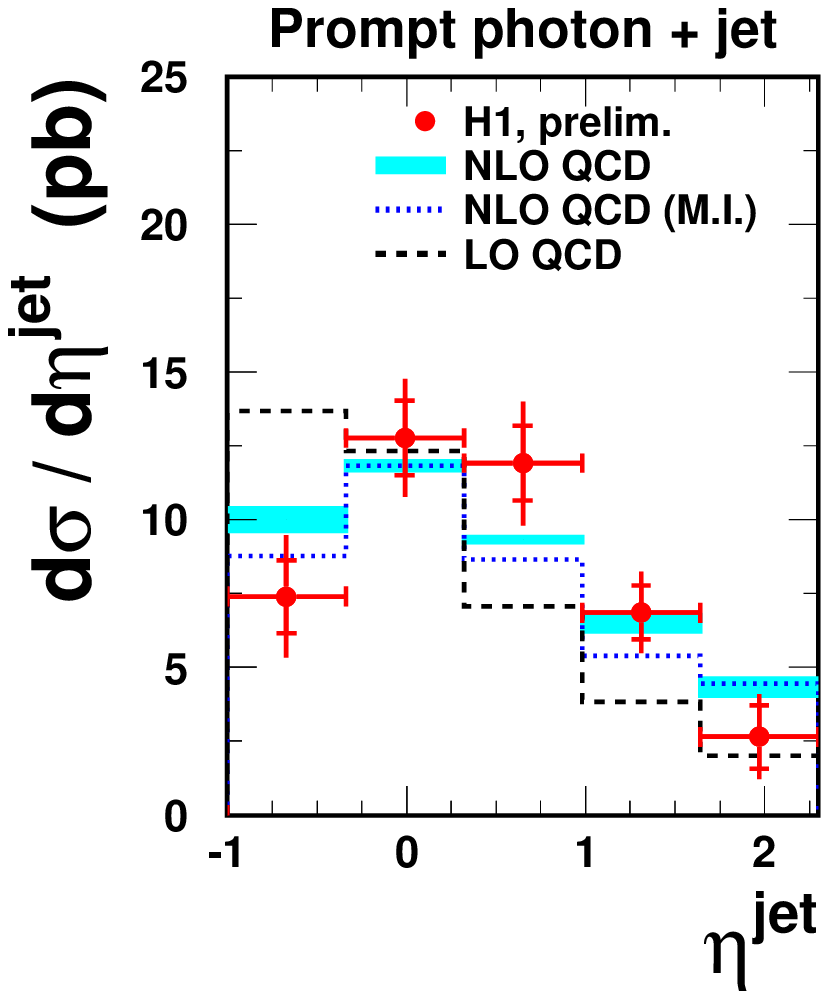}
\includegraphics{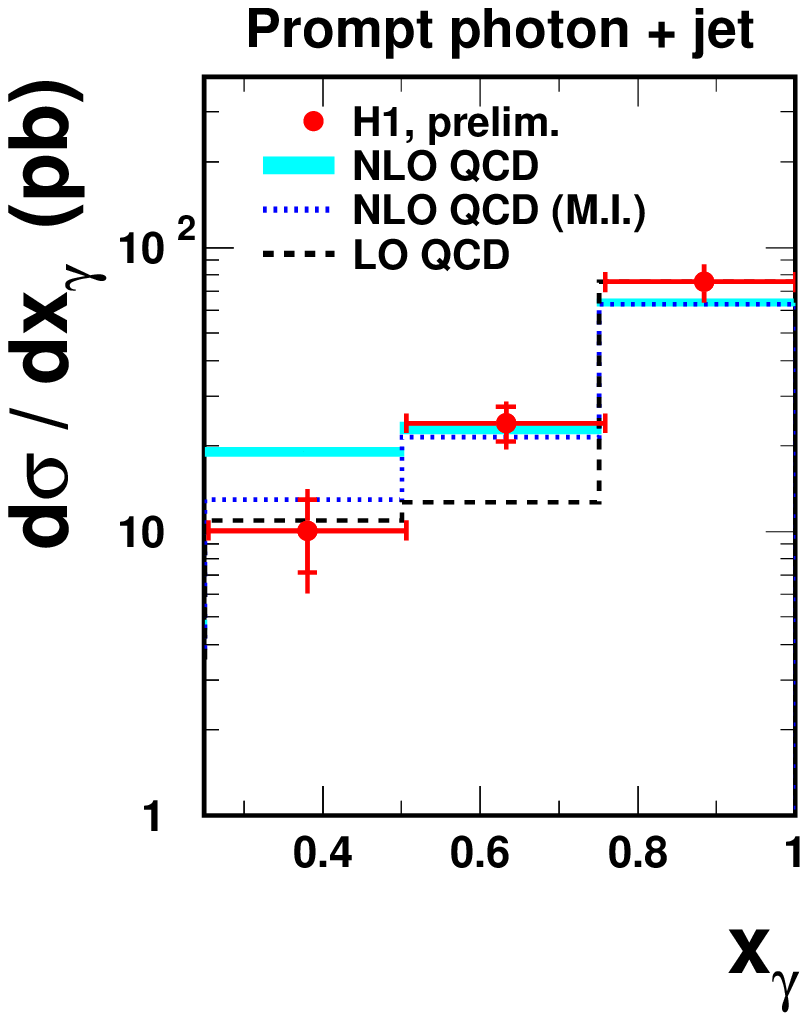}
\includegraphics{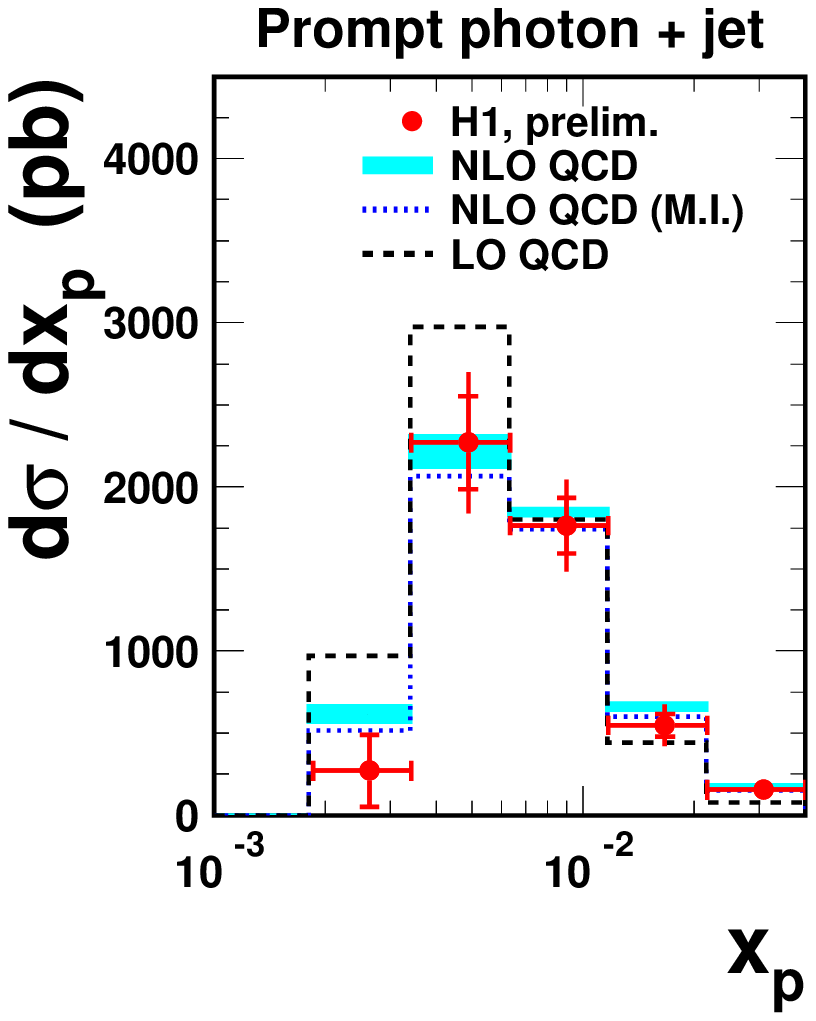}
\caption{
Prompt photon differential cross sections with 
additional jet requirement
  ($E_T^{jet} > 4.5$ GeV)
 as a function of
$E_T^{\gamma}$, $\eta^{\gamma}$, $\eta^{jet}$, $x_{\gamma}$, and $x_p$.
The data are compared with pQCD in LO (dashed line) and NLO
~\cite{Fontannaz:2001ek}.
 The bands show the 
 effect of a variation of the renormalisation and factorisation scales
  from $0.5 \cdot E_T^{\gamma}$
 to $2 \cdot E_T^{\gamma}$. Also shown is the NLO result 
 corrected using the PYTHIA Monte Carlo for
  multiple interaction effects (NLO QCD (M.I.), dotted line).
}
 \label {NLOscale}
\end{center}
\end{figure}

 Cross sections for a prompt photon associated with a jet are presented
 in Fig.~\ref{NLOscale} as function of the variables
    $E_T^{\gamma}, \eta^{\gamma},\eta^{jet}, x_p$ and
 $x_{\gamma}$, where 
 $$x_{p} = (E_T^{jet}e^{\eta^{jet}}+ E_T^{\gamma}e^{\eta^{\gamma}})/2E_p
 \;\;\;\; {\rm and} \;\;\;\;
 x_{\gamma}=(E_T^{jet}e^{-\eta^{jet}}+ 
 E_T^{\gamma}e^{-\eta^{\gamma}})/2yE_e$$

  \hspace*{-18pt} correspond to
  the energy fractions of the incident proton and of the
 exchanged photon participating in the hard process, respectively.

 The data are compared to
    the NLO pQCD calculation
 of Fontannaz et al.~\cite{Fontannaz:2001ek}.
    In the NLO calculation \Etg is used for the renormalisation
 and the factorisation scales.
 The photon and proton parton densities
    AFG~\cite{Aurenche:1994in}
  and MRST2~\cite{Martin:1999ww}
    are used, respectively.
The NLO predictions are substantially different from LO
  and lead to a good description 
of the data. 
 Taking into account the multiple interaction effects, as expected
 on the basis of PYTHIA, improves the data description
 in particular at   $\eta^{\gamma} > 0$ and $x_{\gamma} < 0.5$.
 No corrections have been made to the NLO prediction for hadronisation
 effects.

\section{DIS Results}
\begin{figure}[ht] \unitlength 1pt
\vspace*{256pt}   
\begin{center}
\includegraphics{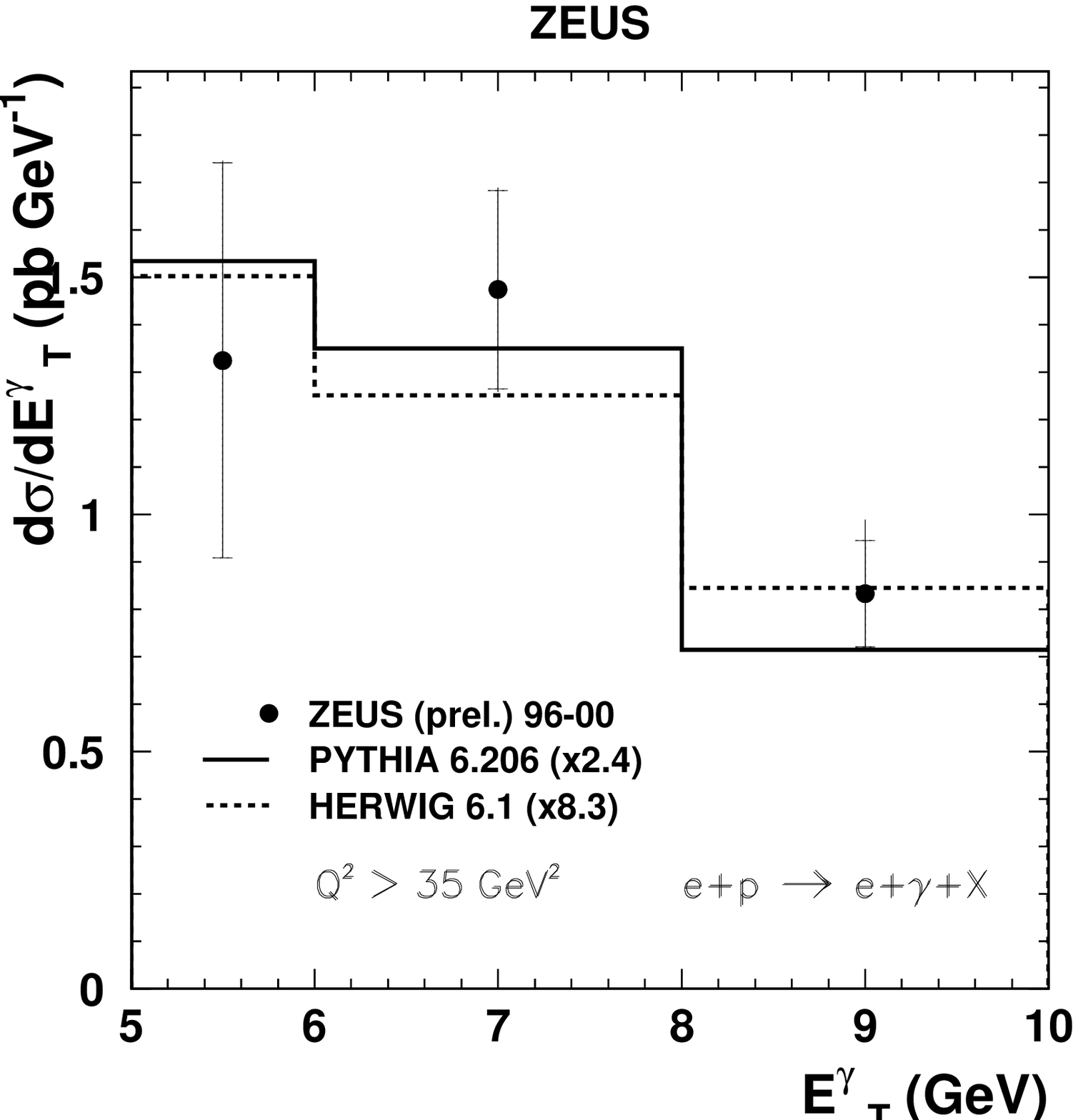}
\includegraphics{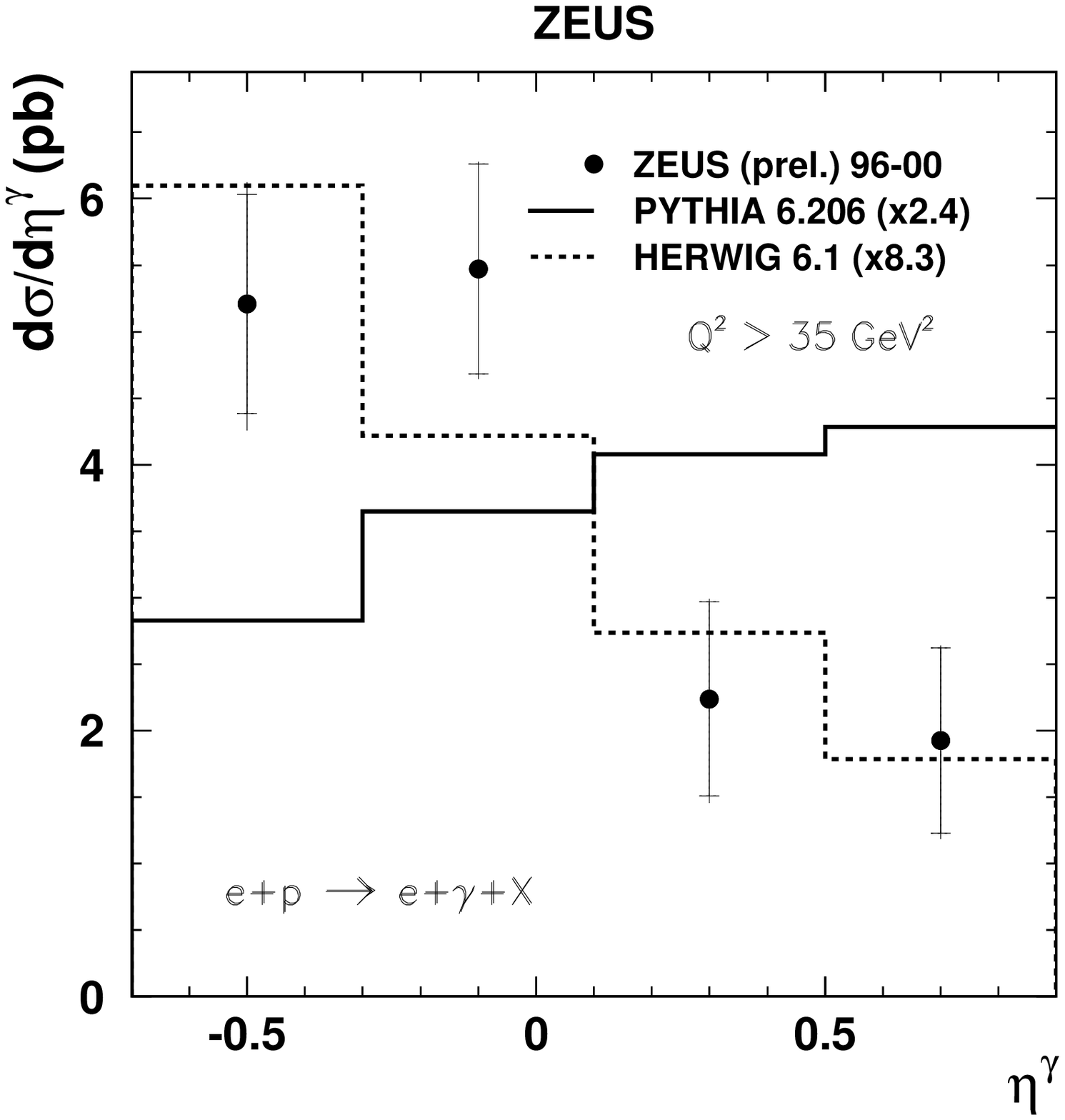}
\includegraphics{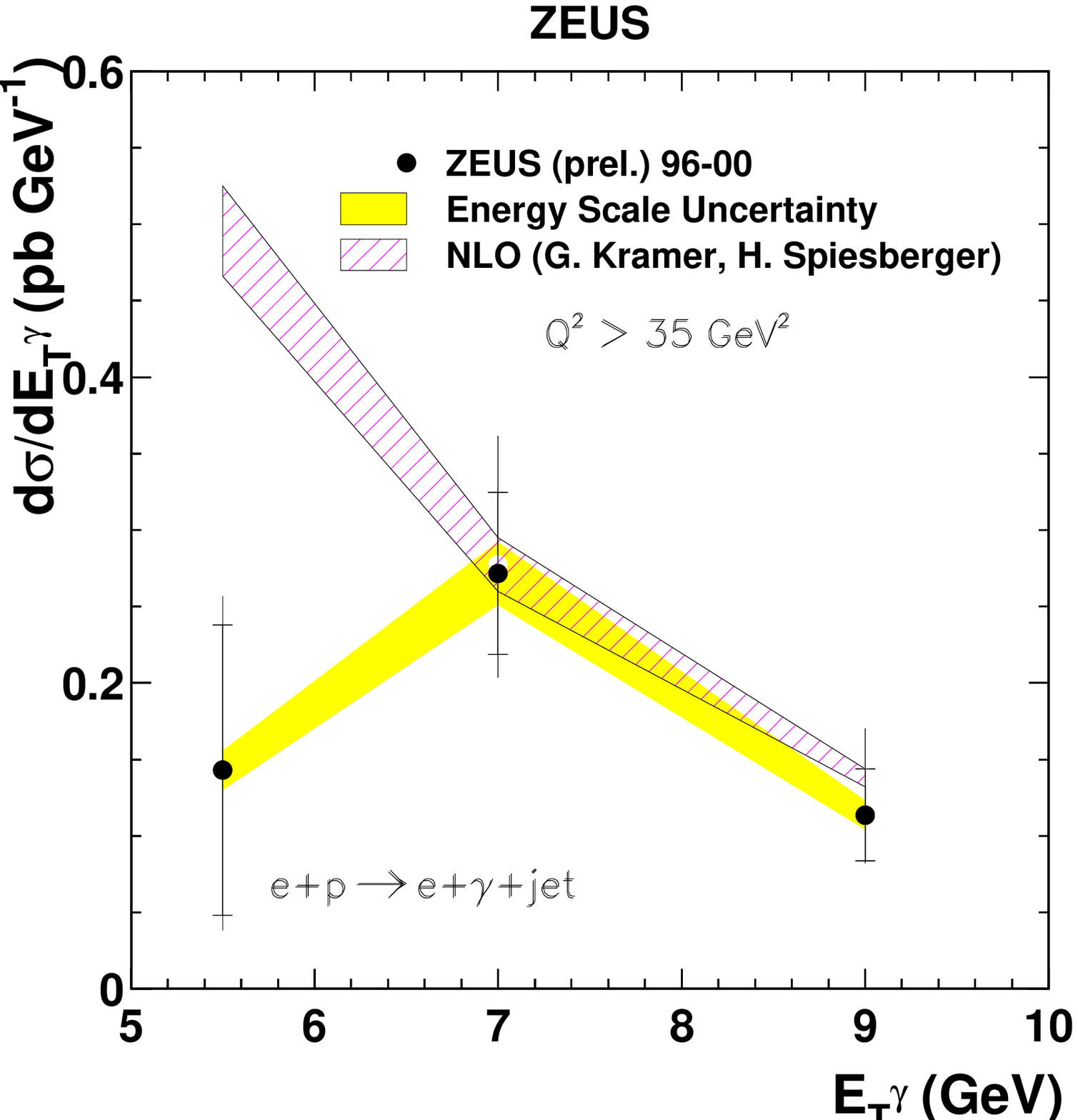}
\includegraphics{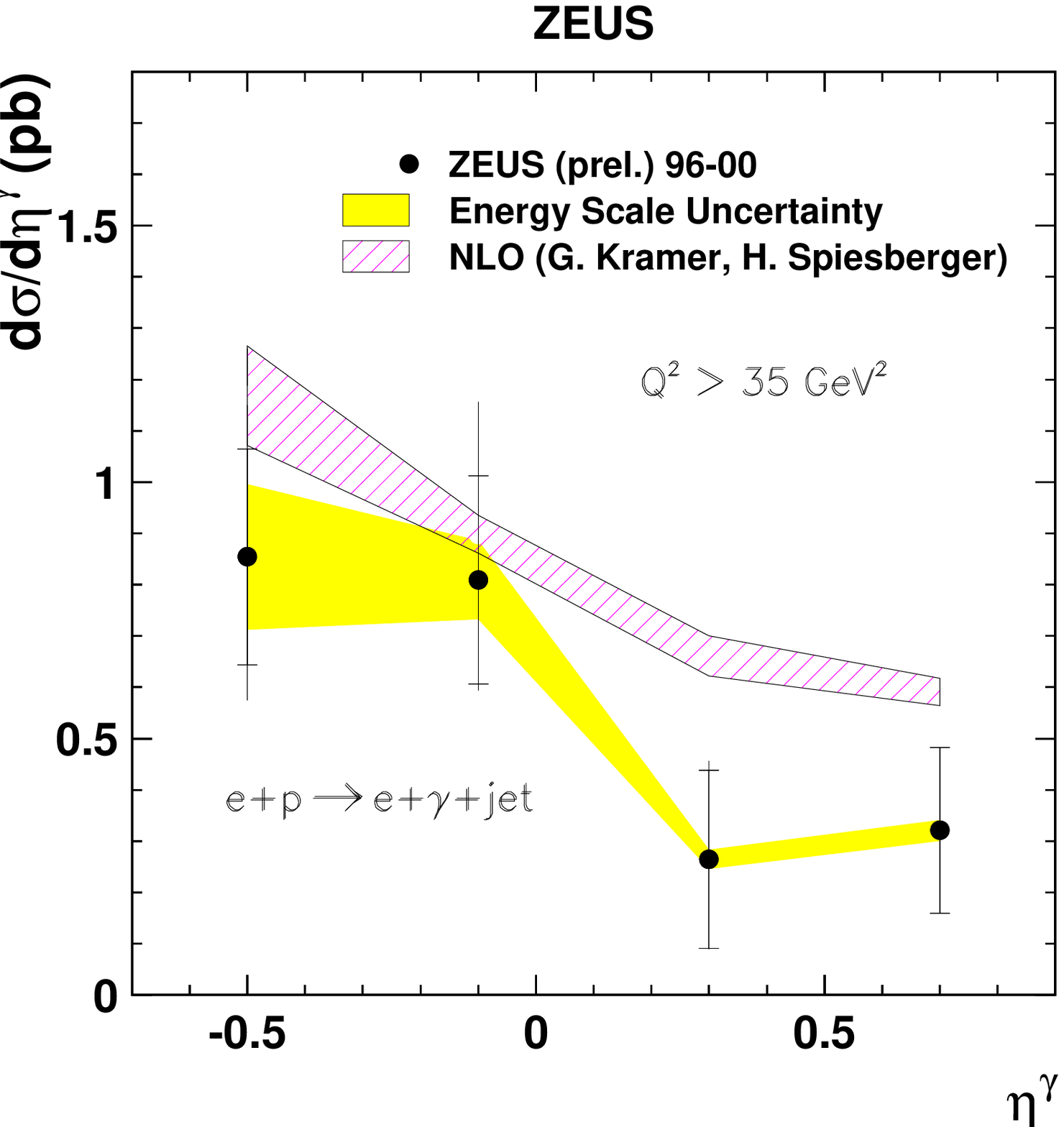}
\caption{
Upper plots: inclusive \Etg and \etag distributions compared with 
 the PYTHIA and HERWIG generators.
Lower plots:  \Etg and \etag distributions with jet requirement
  ($E_T^{jet} > 6$ GeV) compared with pQCD in NLO
~\cite{Gehrmann-DeRidder:2000ce}.
}
 \label {DIS}
\end{center}
\end{figure}
 First results on prompt photon production in DIS are presented by the
 ZEUS collaboration.  
Fig.~\ref{DIS} shows \Etg and \etag distributions for inclusive prompt
photon production (upper row) and associated with one jet (lower row).
After application of normalisation factors of 2.4 and 8.3, respectively,
PYTHIA and HERWIG describe the \Etg distribution, but the \etag distribution
is poorly described by PYTHIA. The pQCD NLO calculation gives an
approximate description of the jet data with no need of an extra
normalisation factor.
    
\section{Conclusions}

 Production of prompt photons has been studied in $\gamma p$ interactions
 and in $ep$ DIS.
 The photoproduction results are quite well described
  by a NLO pQCD calculation, in particular if corrections for multiple
  interactions are applied. 
 Also the DIS data are reasonably well described by a NLO pQCD calculation.
 The HERWIG and PYTHIA models need substantial normalisation factors.


\end{document}